\documentstyle[11pt,moriond,epsfig,axodraw]{article}

\def\be{\begin{equation}}
\def\ee{\end{equation}}
\def\bea{\begin{eqnarray}}
\def\eea{\end{eqnarray}}

\begin{document}

\begin{flushright}
LPT--Orsay 01--46\\
\end{flushright}

\vspace*{4cm}
\title{PHOTOPRODUCTION OF PROMPT PHOTONS AT NLO\,\footnote{Talk given at the 
36th Rencontres de Moriond, "QCD and High Energy Hadronic Interactions", 
Les Arcs, France, March 17--24, 2001}}

\author{G.~HEINRICH}

\address{Laboratoire de Physique Th\'eorique LPT,\\ 
           Universit\'e de Paris XI, B\^atiment 210,\\
           F-91405 Orsay, France}

\maketitle\abstracts{
A numerical program  to calculate the 
photoproduction of prompt photons is presented. 
The code includes full next-to-leading order corrections to all subprocesses.
The results are compared to recent ZEUS data.}

\section{Introduction}

High energy electron--proton scattering 
is dominated by photoproduction processes, where the electron 
acts as a source of quasi-real photons which interact with the partons
in the proton. These processes are of special interest 
since they are sensitive to both the partonic structure of the proton 
as well as of the photon. 
In particular, they offer the possibility  to constrain the 
gluon distribution in the photon, since the subprocess 
$q g\to \gamma q$, where the gluon is stemming from a resolved photon,
is contributing already at leading order.   

The first calculations\,\cite{aur:1984hc}$^-$\,\cite{Gordon:1994sm} 
of higher order corrections to the Compton process 
$\gamma q\to \gamma q$ consider only the fully inclusive cross section 
without the possibility to deal with {\em isolated} photons. 
More recent calculations done by Gordon/Vogelsang~\cite{Gordon:1995km} and 
Krawczyk/Zembrzuski~\cite{Krawczyk} take isolation into account, but 
only by adding a subtraction term evaluated in the collinear approximation 
to the fully inclusive cross section. 
Moreover, the code of~\cite{Krawczyk} does  do not contain  
the full set of NLO corrections. 
   
The calculation presented here takes into account 
the full NLO corrections to direct as well as resolved and fragmentation parts. 
In addition, it includes the box contribution $g\gamma\to g\gamma$ which is
formally an ${\cal O}(\alpha_s^2)$ correction, but known to be important~\cite{Aurenche:1992sb}. 
%The corresponding matrix elements already have been calculated in 
%previous works~\cite{aur:1984hc,ellissexton,Aurenche:1987ff}. 
A major advantage of the present code is also given by the fact that it 
is constructed as a "partonic event generator" and as such is very flexible.
It produces N-tuples of partonic final state configurations which can be 
generated once and for all. Based on these N-tuples, suitable
observables matching a particular experiment can be defined and histogrammed.

\section{Theoretical formalism}

The inclusive cross section for $e p\to\gamma X$ can symbolically 
be written as a convolution of the parton densities of the incident particles
(resp. fragmentation function for an outgoing parton  fragmenting into a
photon) with the partonic cross section $\hat \sigma$
  
\begin{eqnarray}
d\sigma^{ep\to\gamma X}(P_p,P_e,P_{\gamma})&=&\sum_{a,b,c}\int dx_e\int d x_p\int
dz\, F_{a/e}(x_e,M)F_{b/p}(x_p,M_p)\nonumber\\
&&d\hat\sigma^{ab\to c
X}(x_pP_p,x_eP_e,P_{\gamma}/z,\mu,M,M_p,M_F)\,D_{\gamma/c}(z,M_F)
\label{dsigma}
\end{eqnarray}
where $M,M_p$ are the initial state factorization scales, $M_F$ the 
final state factorization scale  and $\mu$ the
renormalization scale. 
The subprocesses contributing 
to the partonic reaction $ab \to c X$  can be divided into four 
categories which will be denoted by
\,1.~direct direct 
\,2.~direct fragmentation 
\,3.~resolved direct 
\,4.~resolved fragmentation.  
The cases 1.~and 3.~correspond to 
$c=\gamma$ and $D_{\gamma/c}(z,M_F)=\delta_{c\gamma}\delta(1-z)$ 
in (\ref{dsigma}), that is, the prompt\,\footnote{By "prompt" we
mean that the photon is not produced from the decay of light mesons.} 
photon is produced directly in the hard subprocess. 
The cases 1.~and 2.~correspond to  $a=\gamma$, with $F_{\gamma/e}$ 
approximated by the Weizs\"acker-Williams formula for the spectrum of 
the quasi-real photons 
\begin{equation}
f^e_{\gamma}(y) = \frac{\alpha_{em}}{2\pi}\left\{\frac{1+(1-y)^2}{y}\,
\ln{\frac{Q^2_{\rm max}(1-y)}{m_e^2y^2}}-\frac{2(1-y)}{y}\right\}\;.
\label{ww}
\end{equation}

For the resolved contributions (3.~and 4.) a parton stemming from the  photon 
instead of the photon itself participates in the hard subprocess, such that 
$F_{a/e}(x_e,M)$ is given by a convolution of the 
Weizs\"acker-Williams spectrum with the parton distributions 
in the photon:

\begin{equation}
F_{a/e}(x_e,M)=\int_0^1 dy \,dx_{\gamma}\,f^e_{\gamma}(y) \,
F_{a/\gamma}(x_{\gamma},M)\,\delta(x_{\gamma}y-x_e)\;.
\end{equation}

At next-to-leading order, the ${\cal O}(\alpha_s)$ corrections to the
corresponding subprocesses are taken into account. The initial state collinear 
singularities are absorbed into the functions 
$F_{a/\gamma}(x_{\gamma},M)$ resp. $F_{b/p}(x_p,M_p)$ 
at the factorization scales $M,M_p$, the final state singularities are absorbed 
into the fragmentation functions $D_{\gamma/c}(z,M_F)$ at the 
fragmentation scale $M_F$. As a consequence, the four subparts separately depend
strongly on $M,M_p,M_F$ and only the sum of all four parts, where the leading 
scale dependence cancels, has a physical meaning. 

From a technical point of view, there
are basically two methods to isolate the infrared singularities 
appearing in the calculation at NLO: The phase space slicing
method and the subtraction method. 
The method used here is basically a phase space slicing method. 
For further details see~\cite{fgh}.

In order to single out the prompt photon events from the huge background
of secondary photons produced by the decays of $\pi^0,\eta,\omega$ mesons,
isolation cuts have to be imposed on the photon signals in the experiment. 
A commonly used isolation criterion is the following: 
A photon is isolated if, inside a cone centered around the photon direction 
in the rapidity and azimuthal angle plane, the amount of deposited 
hadronic transverse energy
$E_T^{had}$  is smaller than some value $E_{T}^{max}$ fixed by the
experiment:
\begin{equation}\label{criterion}
\left(  \eta - \eta_{\gamma} \right)^{2} +  \left(  \phi - \phi_{\gamma} \right)^{2}  
 \leq   R_{\mathrm exp}^{2} \; , \;
E_T^{had}  \leq  E_{T}^{max}
\end{equation}
Following the conventions of the ZEUS collaboration\,\cite{Breitweg:2000su}, 
we used $E_{T}^{max}= 0.1 \,p_T^{\gamma}$  and $R_{\mathrm{exp}}$ = 1.
Isolation not only reduces the background from secondary photons, but also 
substantially  reduces the fragmentation components. 
%In the numerical study described in the next section, the
%fragmentation component is reduced to about 6\% of the total cross section
%after isolation.  

\section{Numerical results and comparison to ZEUS data}

The numerical results can be presented only briefly here, for further details
the reader is referred to~\cite{fgh}. 
For the parton distributions in the
proton the MRST2 parametrization\,\cite{Martin:2000ww} is taken. 
The default choice for the parton distributions in the photon is 
AFG\,\cite{Aurenche:1994in}, 
for  comparisons we also used the GRV\,\cite{Gluck:1992ee} 
distributions transformed to the $\overline{\rm{MS}}$ scheme. 
For the fragmentation functions we use
the parametrization of Bourhis et al\,\cite{Bourhis:1998yu}. 
We take $n_f=4$ flavors and for $\alpha_s(\mu)$ we use an exact 
solution of the two-loop renormalization group
equation, and not an expansion in log$(\mu/\Lambda)$. 
The scales have been set equal to $p_T^{\gamma}$\,; 
a variation of the scales between $p_T^{\gamma}/2$ and
$2p_T^{\gamma}$ leads to a variation of the cross section by less than 10\%. 
The rapidities refer to the $e\,p$ laboratory frame, with  
the proton moving towards positive rapidity. 
To match the ZEUS conventions, we set $Q^2_{\rm{max}}$=1\,GeV 
in the Weizs\"acker-Williams spectrum and restrict the photon 
energy $y=E_{\gamma}/E_e$ to the range $0.2<y<0.9$. 
 
\begin{figure}[htb]
\begin{center}
\mbox{\epsfig{file=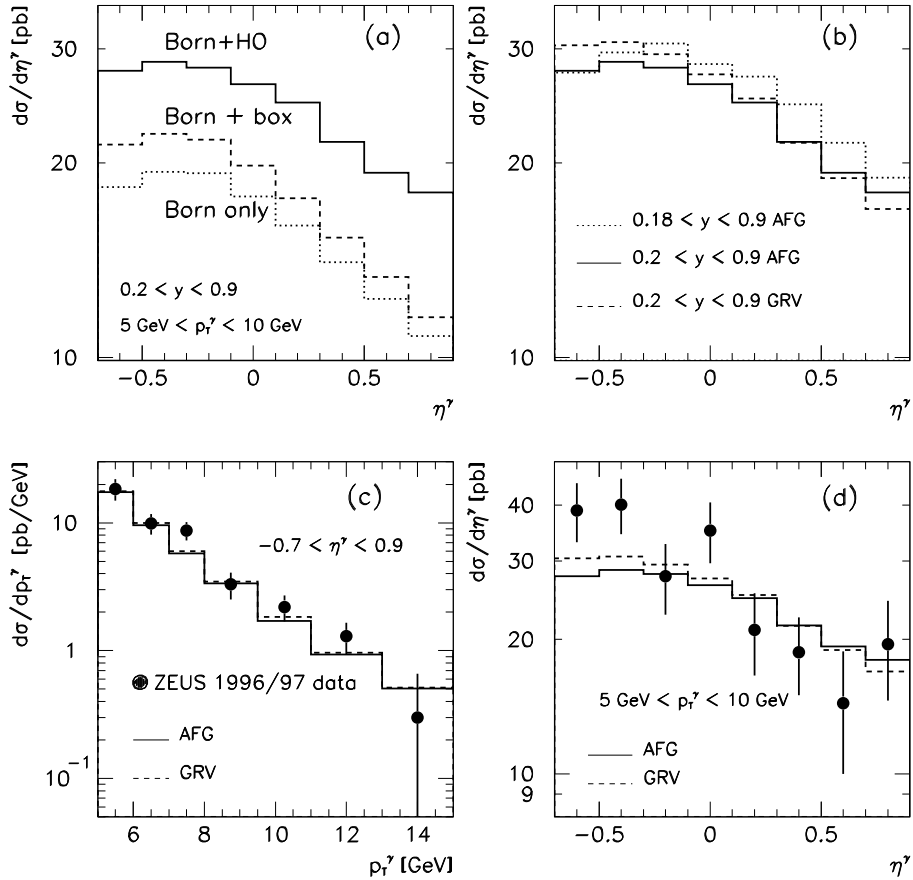,height=13.5cm}}
\caption{(a) Comparison of NLO to LO result for the photon rapidity distribution, 
with isolation. (b) Effect of changing the lower bound on $y$. 
Solid line: default, dotted line: 
lower bound on $y$ decreased by 10\%, 
dashed line: $0.2<y<0.9$ with GRV photon structure functions. (c) and (d):
Comparison to 1996/97 ZEUS data.} 
%for two different sets of parton distributions in the photon.}
\label{all}
\end{center}
\end{figure}

\clearpage

Figure~\ref{all}\,(a) shows a comparison of the NLO 
to the leading order result and displays the magnitude of the box 
contribution. 
The higher order corrections enhance the isolated cross section by about 40\%. 
Fig.~\ref{all}\,(b) illustrates the sensitivity of the cross section 
to a variation of the energy range of the photon: 
A 10\% change of the lower bound on $y$ leads to a spread 
which -- except in the low rapidity region --
 is larger than the one caused by a different set (GRV) 
of photon structure functions. 
Note that experimentally, 
the energy of the incoming photon in photoproduction 
processes is reconstructed from the final hadron energies.  
%with the Jacquet-Blondel method, 
%\begin{equation}
%$y_{JB}=\sum(E-p_z)/2E_e, $
%\label{yjb}
%\end{equation}
In order to obtain the "true" photon energy $y$, corrections for 
detector effects and energy calibration have to be applied. 
%These corrections are assumed to be uniform over the whole $y$ range and 
%enter into the experimental systematic error.  
Fig.~\ref{all}\,(b) shows that a good control on the error in the
reconstruction of $y$ is crucial for a detailed comparison between 
data and theory.  

Figs.~\ref{all}\,(c) and (d) show a comparison to ZEUS 1996/97 
data\,\cite{Breitweg:2000su} for the photon $p_T$ and rapidity distributions 
with two different sets of photon structure functions (AFG and GRV).
For the $p_T$ distribution the agreement is quite satisfactory; 
in the rapidity distribution there is a tendency that theory underpredicts 
the data in the backward region.  

\section{Conclusions}

A full NLO program for the photoproduction of prompt photons has been presented.
The code generates N-tuples of partonic final state configurations which serve
as a starting point to define appropriate observables matching the 
experimental situation. 
%It has been shown that the cross section is very sensitive  to variations of the
%incoming photon energy range. Hence a good control of the experimental error in
%the reconstruction of the photon energy will be crucial for future comparisons.
The agreement between ZEUS 1996/97 data and theory is in general satisfactory; 
a discrepancy can be observed at low photon rapidities. 
With the present errors on
the data, a discrimination between the AFG/GRV sets of parton distributions in
the photon is not possible, but a forthcoming high statistics analysis of all 
1996-2000 data announced by the ZEUS collaboration will improve this situation.

\section*{Acknowledgements}
I would like to thank my collaborators M.~Fontannaz and J.\,Ph.~Guillet.
Further I am grateful to P.~Bussey from the ZEUS collaboration for 
helpful discussions.
%and to the organizers of the conference for the 
%pleasant atmosphere and for financial support.   
This work was supported by the EU TMR Programme, network ''QCD
and the Deep Structure of Elementary Particles'',
contract FMRX--CT98--0194 (DG 12 - MIHT). A special 
grant from the EU to attend the Moriond conference is also gratefully
acknowledged.

%\newpage
\end{document}